# Deep Q-Network Based Resilient Drone Communication: Neutralizing First-Order Markov Jammers

Andrii Grekhov[1*], hrekhov.andrii@npp.kai.edu.ua, ORCID: 0000-0001-7685-8706,
Volodymyr Kharchenko[2], volodymyr.kharchenko@npp.kai.edu.ua, ORCID: 0000-0001-7575-4366
Vasyl Kondratiuk[3], vasyl.kondratiuk@kai.edu.ua, ORCID: 0000-0002-5690-8873

[1*, 2, 3] Aerospace Research and Educational Centre, State University "Kyiv Aviation Institute",
Kyiv, 1, Liubomyra Huzara ave, Kyiv, 03058, Ukraine.

*Corresponding author. [1*] hrekhov.andrii@npp.kai.edu.ua

**Abstract.** Deep Reinforcement Learning (DRL)-based solution for jamming communications using Frequency-Hopping Spread Spectrum (FHSS) technology in a 16-channel radio environment is presented. Deep Q-Network (DQN)-based UAV transmitter continuously selects the next frequency-hopping channel while facing first-order reactive jamming, which uses observed transition statistics to predict and interrupt transmissions. Through self-training, the proposed agent learns a uniform random frequency-hopping policy that effectively neutralizes the predictive advantage of the jamming.

In the presence of Rayleigh fading and additive noise, the impact of forward error correction Bose–Chaudhuri–Hocquenghem (BCH)-type codes (with t = 0–10 correctable errors) is systematically evaluated, demonstrating that even moderate redundancy (t = 1–2) significantly reduces packet loss. Extensive visualization of the learning dynamics, channel utilization distribution, ε-greedy decay, cumulative reward, BER/SNR evolution, and detailed packet loss tables confirms convergence to a near-optimal jamming strategy. The results provide a practical framework for autonomous resilient communications in modern electronic warfare scenarios.

## Introduction

The rapid development of Unmanned Aerial Vehicles (UAVs) for civilian and military use has dramatically increased the demands on the reliability and jamming immunity of communication channels in active electronic warfare environments. Modern reactive (tracking) jammers are capable of detecting transmissions and instantly concentrating power on the occupied channel. Among these, first-order Markov reactive jammers pose a particular threat: by observing the hopping sequence, they construct a transition probability matrix in real time and increasingly accurately predict the transmitter's next channel.

Classical Frequency-Hopping Spread Spectrum (FHSS) systems using pseudorandom sequences (m-sequences, cryptographic generators) and even hardware sources of truly random numbers are fundamentally vulnerable to such a learning adversary and, under typical conditions (jamming/signal ratio ≈ 1–2), do not ensure the successful transmission of all packets. Recent advances in Deep Reinforcement Learning (DRL) have opened up entirely new possibilities for creating fully autonomous, jam-resistant communication systems. Unlike traditional approaches based on predetermined hopping sequences, a DRL-based transmitter can autonomously learn an adaptive strategy directly while interacting with a jamming environment, without any a priori knowledge of the adversary model.

Despite the emergence of DRL-based solutions, most employ complex architectures (Double DQN, Dueling DQN, recurrent DRQN, multi-agent games), which significantly increases computational complexity and makes implementation onboard small UAVs extremely challenging.

This paper examines whether a lightweight classical Deep Q-Network (vanilla DQN) with one-hot encoding of the current channel and minimal modification of the reward function can generate an optimal hopping strategy that completely neutralizes a first-order Markov reactive jammer.

We consider a slow 16-channel FHSS system operating under Rayleigh fading and additive white Gaussian noise, attacked by a reactive jammer with interference-to-signal ratios of 0.5 and 1.2. The jammer observes previous transitions and builds an accurate first-order Markov model in real time. The transmitter has no information about the jamming strategy and learns solely by trial and error.

The objective of the study is to:

- demonstrate that a conventional single-agent DQN with a small diversity bonus converges to a nearly uniform channel selection strategy,

- quantify the achieved transmission success rate, Bit Error Rate (BER), Signal-to-Noise Ratio (SNR), and Packet Loss Probability (PLR) with and without Bose–Chaudhuri–Hocquenghem (BCH) coding under a given interference environment.

The rest of the paper is organized as follows. Section 2 covers related work. Section 3 presents DQN mathematical model for FHSS and reactive jammer. In Section 4, description of algorithm is given. Results of simulation are considered in Section 5. Discussion is presented in Section 6. Conclusions are given at the end of the article.

## 1. Related Works

In [1], a new resource allocation strategy for interference mitigation in cognitive radio is proposed, using a cognitive UAV as an example. An active Generalized Dynamic Bayesian Network (Active-GDBN) is proposed as an environment model, which jointly encodes the physical signal dynamics and the dynamic interaction between the UAV and the spectrum jammer. Actions and planning are considered as a Bayesian inference problem, which can be solved by avoiding unexpected states (anomaly minimization) in the online learning process. Simulation results confirm the effectiveness of the proposed approach in anomaly minimization (reward maximization) and demonstrate high convergence rate compared to traditional frequency hopping and Q-learning methods.

In paper [2], a reinforcement learning-based UAV relay policy for maritime communications is proposed to counter jamming. Based on previous transmission characteristics, relay location, received transmitted signal strength, and received jamming power, this scheme optimizes the UAV trajectory and relay power to save energy and reduce the BER of maritime signals. A deep reinforcement-learning scheme is also proposed, which designs a deep neural network with a dueling architecture to improve communication quality and computational complexity. Performance bounds on the signal-to-interference plus noise ratio, energy consumption, and communication utility are determined based on the Nash equilibrium of the anti-jamming game, and the computational complexity of the proposed schemes is analyzed. Simulation results show that the proposed schemes improve energy efficiency and reduce BER compared to the benchmark.

Paper [3] presents a new approach JaX for detecting and suppressing powerful jammers in situations where traditional spread spectrum and other jamming mitigation methods are insufficient. JaX requires no explicit soundings, training sequences, channel estimation, or transmitter interaction. A convolutional neural network is developed for a multi-antenna system to detect the presence of jammers, the number of interfering emissions, and their corresponding phases. This information is continuously fed to an algorithm that suppresses the interfering signal. A prototype system with two antennas is developed, and the approach is evaluated under various environmental conditions and modulation schemes using software-defined radio platforms. JaX is robust to various jammer types with varying jamming signal characteristics, jamming power, and timing patterns.

UAV communication networks are vulnerable to malicious jamming and co-channel interference, which degrade network performance. Therefore, researching anti-jamming methods to improve communication security has become a serious challenge. In paper [4], a new anti-jamming channel selection scheme is proposed in a multi-channel network with multiple UAVs. First, the anti-jamming problem is formulated as Partially Observable Stochastic Game (POSG), in which pairs of UAVs with partial observability compete for a limited number of communication channels with a Markov jammer. To ensure fast adaptation to the dynamic jamming environment, Meta-Mean Field Quality (MMFQ) learning algorithm is proposed, which provides Nash Equilibrium (NE)-based solution to the POSG problem. Simulation results show that the proposed algorithm can achieve a higher average result than benchmark algorithms, contributing to improved throughput and increased resource utilization, especially for large-scale UAV communication networks.

With the growing popularity of UAVs in both military and civilian applications, it is important to ensure the reliability and security of their signals. In recent years, deep reinforcement learning has proven itself as a powerful tool for addressing this challenge. The purpose of paper [5] is to analyze the potential of deep learning in combating UAV communication system jamming. Traditional jamming countermeasures, such as frequency hopping and Direct Successive Spread Spectrum (DSSS), are effective to a certain extent. However, they are challenging to cope with dynamically changing jamming environments due to their single-mode nature and low spectrum utilization. On the other hand, deep learning has powerful capabilities in automatic feature extraction and pattern recognition, especially deep reinforcement learning, which enables UAVs to adapt to environmental changes in real time and automatically optimize communication strategies. This article examines anti-jamming methods based on reinforcement learning and deep reinforcement learning algorithms, and proposes a new anti-jamming strategy for UAV communications based on Generative Adversarial Networks (GANs). Although deep learning offers new opportunities for combating interference in UAV communication signals, it still faces challenges related to high computational demands and complex model training in practical applications.

The objective of review [6] is to summarize a small number of narrowband interference (NBI) models and mitigation methods described in the literature over the past decade for various commercial and defense domains, including terrestrial and satellite systems. The sources and potential negative impacts of NBI on various applications and technologies are discussed. A range of mitigation strategies are considered, including traditional filter-based methods and modern methods using machine learning. The review is limited to time-frequency NBI and does not consider the interference problem in the spatial domain, given the number of publications published to date. Some identified research gaps are discussed and future research directions for NBI mitigation are listed. This review introduces the NBI problem and can be used in future research on emerging applications and technologies.

Jamming attack is a cyber-threat that leads to denial of service, which often occurs in wireless communication systems such as Flying Wireless Networks (FANETs) and the Internet of Drones (IoD). Over the years, several approaches have been proposed for jamming attack detection, such as the Bayesian game theory mechanism, the IoD-based defense mechanism, communication channel methods (channel hopping, spread spectrum, MIMO-based interference suppression, coding, etc.), the delay-tolerant network method, and cryptographic algorithms. However, these methods are not suitable for jamming detection in the UAV environment. The main problems relate to delivery efficiency, processing time, accuracy, power consumption, flight range, and flight autonomy. In [7], a jamming attack detection method is presented using the reinforcement learning mechanism based on gradient monitoring (RLGM). RLGM maintains safe regions and reduces the gradient variance for the intended learning, which ensures higher accuracy of the learning target. RLGM achieves rapid learning progress and selects precisely the set of parameters required by the network during the training phase. RLGM spontaneously infers the required deep network scale during training, using automatically immutable learned weights. The proposed approach outperforms other reinforcement learning methods, such as federated reinforcement learning, Deep Quadrature Learning (DQL), and non-machine learning methods such as GA-AOMDV.

In [8], a novel intelligent jamming countermeasure framework is presented for UAV networks. Multiple UAV-UAV link pairs strive to maximize their combined rates while minimizing power consumption, with each UAV adaptively adjusting its transmit channel and allocated power to avoid intelligent jamming and interference on the shared channel. A jammer attempts to disrupt the UAV network's link quality by adaptively changing its jamming channel and power. The jamming countermeasure problem is modeled as a stochastic Stackelberg game, where the intelligent jammer is the leader and the UAV pair are followers. Reinforcement learning (RL) algorithms are proposed to determine the best response policy of each agent in the game. We use a deep Q network (DQN) algorithm to detect jamming at stations and propose a decentralized federated DQN algorithm with detection support for suppressing joint policy jamming in UAV pairs. The analysis results show that the anti-jamming efficiency of the proposed algorithm is 23.3% compared with the independent DQN algorithm.

Spectrum scarcity, spectrum efficiency, power limitations, and jamming are the main challenges facing wireless networks. Cognitive Radio Networks (CRNs) enable the sharing of licensed bands when they are available. Secondary Users (SUs) to ensure high data rates must efficiently use spectrum, and SU mobility makes power consumption an important factor in wireless networks. Due to the open environment, jamming can easily degrade performance and disrupt connections. A study [9] aims to improve CRN performance and establish more reliable connections for SUs in the presence of intelligent jammers, ensuring efficient spectrum use. An approach to combat jamming using frequency hopping is proposed. SUs are assumed to monitor spectrum availability and channel gain. The SUs then learn the jammer's behavior and select an appropriate policy regarding the number of data and control channels, which jointly optimizes spectrum efficiency and power consumption. The interaction between the SU and the jammer is modeled as a stochastic zero-sum game, and Reinforcement Learning (RL) is applied to solve this game. Simulation results show that low channel gain leads the SU to select a large number of data channels. However, when the channel gain is high, the SU increases the number of control channels to ensure a more reliable connection. Considering spectrum efficiency, the SU saves energy by reducing the number of channels used. The proposed strategy achieves better performance compared to short-sighted learning and a random strategy. Under jamming attacks, the SU selects an appropriate number of control and data channels to ensure a reliable, efficient, and long-term connection.

UAV communication systems face increasingly serious challenges related to multi-source jamming in dynamic countermeasure environments, placing increased demands on their reliability and robustness. To address these issues, agent-based autonomous jamming methods have become a key research area. Paper [10] presents a comprehensive review that formalizes the concept of intelligent agents for jamming UAV communications and proposes a closed-loop decision-making system based on the Perception-Decision-Action (P-D-A) paradigm. Within this framework, key technologies are examined, with a particular emphasis on the application of game theory to modeling UAV interactions with jammers and the integration of intelligent reinforcement learning algorithms to develop adaptive jamming strategies. Potential limitations of existing approaches are discussed, critical engineering challenges are identified, and promising directions for future research are outlined.

The paper [11] provides an overview of the evolution of intelligent interference suppression methods in communications. The concept is defined and the main characteristics of the capabilities of intelligent interference suppression in communications are discussed. The initial architecture structure of an intelligent interference suppression system in communications is described. The development of intelligent interference suppression methods in communications is considered, tracing its development from early adaptive interference suppression methods to more modern advances in the field of intelligent interference suppression, based mainly on game theory and machine learning. The latest research results in this area are analyzed and existing problems are highlighted. Several promising directions for future research in the field of intelligent interference suppression are proposed.

In [12], the threats posed by highly dynamic intelligent jamming are addressed and an intelligent jamming algorithm based on proximal policy optimization (PPO) is proposed. The problem of communication noise immunity under highly dynamic jamming is modeled as a multi-parameter Markov decision process (MDP). Then, the proximal policy optimization algorithm is used to determine the optimal joint jamming strategy for the system channel, time interval, transmission rate, and transmission power. This approach improves the convergence rate of the algorithm and reduces the time required to reach the optimal strategy. Simulation results show that the proposed algorithm effectively adjusts multiple communication parameters simultaneously, demonstrating superior statistical performance in packet reception ratio and convergence rate compared to existing jamming methods based on Q-learning and DQN.

Deep reinforcement learning is widely applied to solving anti-jamming problems in wireless communication. However, it is assumed that the communication system can obtain complete channel state information (CSI). Under limited CSI, the system is modeled in [13] using partially observable Markov decision processes. This paper proposes an algorithm for automatically adjusting the search rate rolloff factor. Furthermore, a deep recurrent Q-network algorithm architecture and an intelligent anti-jamming decision algorithm are developed. The algorithm first uses long short-term memory networks to learn the temporal characteristics of the input data, equalizes the characteristics, and then feeds the result into fully connected layers to obtain an intelligent anti-jamming strategy. Simulation results show that the automatic search rate rolloff factor-tuning algorithm can achieve near-optimal performance when setting a large initial search rate rolloff factor.

The paper [14] proposes a multi-agent approach to spectrum access control based on deep reinforcement learning (DRL) and a value decomposition network (VDN) with centralized training and a distributed execution architecture. Following centralized ground-based training, the model was deployed autonomously across satellites to make spectrum access decisions in real time. Simulation results show that the proposed method effectively balances training costs with jamming mitigation effectiveness.

Report [15] examines the problem of protecting UAV communication channels from adaptive AI jamming. A countermeasure model is proposed in which the defensive AI onboard UAV uses Q-Learning for dynamic frequency switching, while the attacking AI employs a transition prediction model. A simulation is conducted and it is shown that the RL agent ensures stable communication even with aggressive attacker adaptation.

## 2. DQN Mathematical Model for FHSS vs. Reactive Jammer

A *mathematical model* of the problem is developed for a single-agent DQN in FHSS scenario, with Markov jammer, Rayleigh fading, and AWGN channel (Fig. 1). The formulas below form the key block, focusing on SNR/BER, reward, entropy, and success rate.

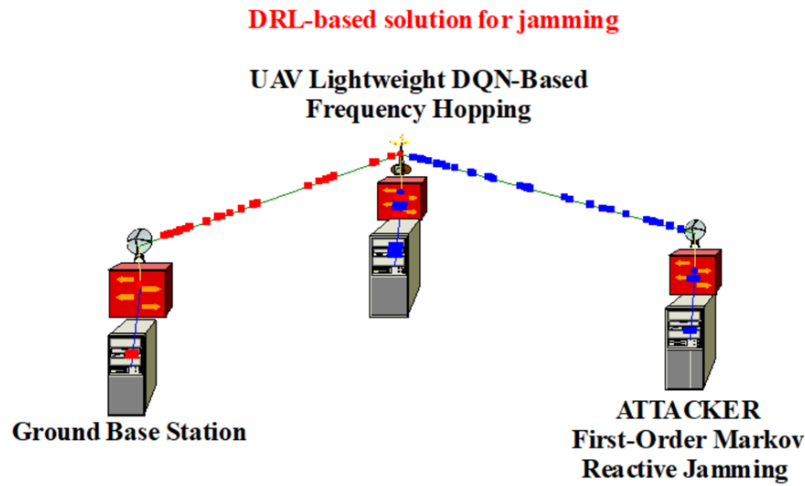

**Fig. 1**. Model "Ground Base Station –UAV – ATTACKER"

The *scenario* is that DQN agent manages FHSS with N=16 channels for transmitting packets in the contested spectrum. The jammer is reactive, first-order Markov, and switches to the agent's previous channel with probability 0.8. The hop is slow (one per packet). The channel is Rayleigh fading, and Jammer-to-Signal Ratio (JSR) varies from 0.5 to 1.2. Success is defined as no hop and jam overlap.

The *parameter t* is the time step in a discrete time model. It is an index denoting a point in time in a sequence of events. In the model, the agent interacts with the environment at each step: it observes a state, chooses an action, receives a reward, and moves on to the next state. The model is discrete, where time is broken down into "ticks" to simulate sequential decisions—channel switching. The model has $N = 16$ channels (numbered 1 through 16). The current state is a one-hot vector of length 16, where 1 is the current channel position and the rest are zeros.

The *action* $a_t \sim \pi(s_t)$ is the choice of the next channel according to policy $\pi$ (a probability distribution). The jammer is modeled as a first-order Markov chain: transition P(jam $c_{t+1}|c_t$) $\propto$ count($c_t \rightarrow c_{t+1}$), meaning the probability of jamming at $c_{t+1}$ depends on the transition rate from $c_t$. Ultimately, the agent tries to choose $c_{t+1}$ to avoid the jammer, but the jamming is time-correlated.

The *parameter* $|h|^2$ is the squared absolute value of the channel coefficient $h$, which represents the channel gain at a given step. In wireless communications, $h$ is a complex attenuation coefficient (fading), including amplitude and phase. The absolute value of $h$ is the amplitude, and $|h|^2$ is the instantaneous signal power after passing through the channel.

In the formula for the *Signal-to-Noise Ratio* $\gamma = SNR = \frac{P_s \cdot |h|^2}{P_n + P_j \cdot |h|^2 \cdot I_{jammed}}$, $P_s$ is the signal power, $P_n$ is the noise, and $P_j$ is the jammer power. The parameter $I_{jammed} = 1$ if the channel is jammed. The $|h|^2$ parameter takes into account how the channel "boosts" or "attenuates" the signal. If $|h|^2$ is small, the SNR drops, which degrades the connection quality.

The *channel coefficient h* has a distribution according to Rayleigh's law, the standard model of Rayleigh fading in wireless channels. The physical meaning is that in channels without line-of-sight (Non-Line-of-Sight, NLOS), the signal arrives along many reflected paths (multipath propagation). The amplitude $h$ is the sum of these paths, which behaves as a random variable with a Rayleigh distribution. This models the fading due to beam interference. The parameter $h$ is complex: $h = X + iY$, where $X$ and $Y$ are independent Gaussian random variables with zero mean and variance $\sigma^2$. Then the amplitude $r = |h| = sqrt\{X^2 + Y^2\}$ has a Rayleigh distribution. The power g $= |h|^2 = X^2 + Y^2$ has an exponential distribution. Fading may leads to "deep fades" when $|h|^2$ is close to 0 and the connection is broken.

In the model, $h$ is generated anew at each step t (independently), so $|h|^2$ is a random power. The SNR becomes random, and the *Bit Error Rate* (BER) (for BPSK modulation with coherent reception) is:

$$BER(t) = Q\left(\sqrt{2\gamma(t)}\right) = \frac{1}{2}erfc(\sqrt{SNR(t)}\ ),$$

where *Q(x)* is the tail function of the Gaussian distribution $erfc(x) = \frac{2}{\sqrt{\pi}}\int_x^\infty e^{-u^2}\,du$ .

*Packet loss* is estimated using the following formula:

$$PLR = 1 - (1 - BER_k)^L \approx L \cdot BER_k \text{ for small } BER.$$

### DQN agent model

• State: $s_t$ = (prev_channel, jammer_history) – one-hot for 16 channels + Markov jammer (dimension 16 +1 = 17).

• Action: $a_t \in \{1, ..., 16\}$ – choosing a channel.

• Q-function: tabular DQN *Q(s, a; θ)*, update:

$$\theta \leftarrow \theta + \alpha[r + \gamma \max_{a'} Q(s', a';\ \theta) - Q(s,a;\theta)]\nabla_\theta Q(s,a;\theta)$$

• Policy distribution: Softmax with temperature $\tau = 1$ (for exploration):

$$\pi(a|s) = \frac{\exp(Q(s,\ a)/\tau)}{\sum_{a'} \exp(Q(s,\ a')/\tau)}$$

In the context of our model (RL for channel selection), Softmax is a key tool for transforming estimates (logits) into a probability distribution over actions (channels). This is especially relevant for $\pi(s_t)$ policy, where the agent chooses the next channel $c_{t+1}$ stochastically to balance exploitation (selecting the best channels) and exploration (trying new ones).

*Softmax* (or "soft maximum") is a function that takes a vector of numbers (logits, such as the Q-score) and transforms them into a probability distribution where the sum of the probabilities is 1 and all values are ≥ 0.

It is used in the RL model because the agent does not always strictly choose the best channel to avoid getting stuck. Softmax makes the choice "soft": the best channels are chosen more often, but the worst ones are chosen with low probability. This encourages exploration.

In our model, for $N = 16$ channels, the logit vector z = [$z_1, z_2, ..., z_{16}$], where $z_i$ is the valuation for channel $i$ (e.g., the expected reward). Softmax yields $p_i = \pi(a_t = i \mid s_t)$,, and $a_t$ is sampled from this $p$.

*Basic Softmax formula* (for $\tau = 1$):

$$p_i = \frac{\exp(z_i)}{\sum_{j=1}^N \exp(z_j)} \ , i = 1, ..., N.$$

*Softmax with temperature* $\tau > 0$ (temperature scaling), which is added to control exploration:

$$p_i = \frac{\exp(z_i\ /\ \tau)}{\sum_{j=1}^N \exp(z_j/\ \tau)}$$

The *role of $\tau$* is as follows:
• $\tau = 1$: standard Softmax, where the balance is that differences are preserved but there is exploration (best actions are more frequent, but not 100%).
• $\tau > 1$: hot temperature, a "hot" policy where dividing by $\tau$ "smooths" the exponents and the distribution becomes more uniform. More exploration, where the agent more often tries bad channels to find hidden patterns in the jammer.
• $\tau < 1$: cold temperature, where differences are amplified. More exploitation, where the agent persistently chooses top channels, ignoring risks.

*In our case, $\tau = 1$ for exploration*, which is the baseline learning mode. It is stochastic enough to prevent the agent from becoming "stuck" on a single channel (given the Markov jammer), but not too chaotic. This is often the starting point, and $\tau$ can later be reduced to converge to the optimal policy.

Our model uses a *reward function* within a reinforcement-learning framework so that the agent (the channel selection system) learns to maximize the cumulative reward per episode. This motivates avoidance of interference (jammers), ensuring good connection quality (low BER), and encouraging exploration. The formula is:

$$r_i = -2 \cdot I\{jammed\} + \frac{1}{1 + BER} + \beta \cdot H(p),$$

where:
• $I\{jammed\}$ – indicator (1 if the channel is jammed at step $t$, otherwise 0).
• $BER$ – bit error rate, from 0 (perfect) to 1 (complete chaos).
• $p$ – probability distribution over channels (from policy $\pi$).
• $H(p)$ – Shannon entropy.
• $\beta = 0.02$ – coefficient for the entropy bonus.

The *reward* is calculated at each time step when the agent selects a channel, transmits data, and receives feedback (jammed, SNR, BER). The base reward is calculated using the formula $\frac{1}{1+BER}$ and reflects the quality of data transmission on the channel. When the BER is low (good channel), the reward is close to 1; when it is high (bad channel), it is close to 0.5.

This formula was chosen because it is a monotonically decreasing function (the higher the BER, the lower the reward), but not linear—it is "soft" at low BERs (small errors do not ruin everything) and steep at high BERs (channel failure heavily penalizes). This is typical for communications, where BER is exponentially dependent on SNR.

The *base reward* encourages the choice of channels with high SNR (low fading $|h|^2$ and no jammer). Without this, the agent would ignore the connection quality, focusing only on avoiding the jammer. The base reward scale is usually 0.5–1, to be the "main" part of the reward (other terms adjust it).

The *jamming penalty* is $-2 \cdot I$. Jamming is artificial interference from a jammer that adds power $P_j$ to the SNR denominator, making the channel practically unusable (SNR $\rightarrow$ 0, BER $\rightarrow$ 0.5). The penalty is a negative bonus that "penalizes" the agent for choosing a jammed channel: $I\{jammed\} = 1$ if the jammer is active on $c_{t+1}$, otherwise 0.

The value -2 is chosen as an asymmetric penalizer. Since the base reward is max = 1 (clean channel), the value -2 makes jamming "very bad" ($r \approx -1$), so the agent urgently learns to avoid it. If the penalty were -0.5, the agent could "tolerate" jamming occasionally, but the value -2 creates a strong gradient in learning.

The *logic* here is as follows. In real networks, a jam is a disaster (packet loss, delay). A penalty of 2 times the maximum base reward ensures that avoiding the jammer is the number one priority (exploration/exploitation balance). The indicator $I$ takes a value of 1 or -2 and "triggers" the penalty only during a real jam. This is binary: there are no grayscale values—either there is a jam (a complete failure) or there is not. The model uses a Markov jammer, and thus the agent learns to predict transitions. This creates a negative signal for bad behavior. Without the penalty, the agent could ignore the jam, relying solely on the BER (which is high, but not extreme, during a jam).

The *entropy bonus $\beta \cdot H(p)$* is an additional term to encourage exploration. The parameter $H(p) = -\sum_{i=1}^{N} p_i \log p_i$ is the Shannon entropy of the policy distribution $p$ (Softmax over 16 channels). The entropy measures the "randomness" of the choice. The parameter $\beta = 0.02$ provides scaling. An entropy bonus is introduced because pure exploitation in RL leads to local optima (the agent gets stuck on 1–2 channels, unaware of jammer shifts). Entropy adds "noise" to the reward, which stabilizes learning. The value $\beta = 0.02$ was chosen empirically to avoid outweighing the base reward. The role of the parameter $\beta$ is to balance the short-term consequences (avoid jamming now) and the long-term consequences (adapt to the Markov jammer).

*Success rate* is determined by the formula: $Success = \frac{1}{T}\sum_{t=1}^{T}\left(1 - I_{a_t}\right) \times 100\%$, where T = 1500 episodes.

### Markov chain-based reactive jammer model

In the context of our model, a reactive jammer is an adaptive antagonist that reacts to the agent's actions by switching between channels to create interference. Unlike a "dumb" jammer (random or fixed), a reactive model learns from the agent's previous transitions to predict and block its choices. This is implemented as a first-order Markov chain, where the jammer state is the currently jammed channel, and transitions depend on the agent's action history.

The model makes the jammer "smart": it is not random but time-correlated, which increases the challenge for the RL agent (it must learn to predict shifts).

*The concept of a reactive Markov jammer*

*Reactivity*: The jammer "reacts" to the agent by monitoring its transitions between channels. It assumes that the agent has patterns (e.g., preferring adjacent channels) and adapts its transitions accordingly. This models real-world scenarios.

*First-order Markov property*: The current state (jam on channel $c_t$) determines the probability of the next state (jam on $c_{t+1}$) only through the agent's previous state. There is no dependence on earlier history—this is a simplification for tractability (computational simplicity).

*Key element*: count($c_t \rightarrow c_{t+1}$): The jammer keeps a "counter" of the agent's transition frequency. The more often the agent transitions from $c_t$ to $c_{t+1}$, the higher the probability that the jammer will "transition" there to block. This makes the jammer evolvable: at the beginning of the simulation (RL episode), the transitions are almost random; as the agent learns, the jammer "learns" from its policy $\pi$.

*Jammer state*: $j_t \in \{1, ..., N\}$ — the channel jammed at step $t$.

*Jammer action*: Automatic — sampling $j_{t+1} \sim \mathrm{P}(j_{t+1} \mid c_t)$, where $c_t$ is the agent's last choice. It does not suppress all channels at once (too energy consuming), but focuses on the agent's "hot" paths, minimizing its energy.

*Mathematical model*

*States*: Discrete channels $\{1, 2, ..., 16\}$. The total state space is $N = 16$.

*Transition matrix*: $P$ is a stochastic matrix of size $N \times N$, where $P_{i,k} = \mathrm{P}(j_{t+1} = k \mid c_t = i)$ is the probability of the jammer transitioning to channel $k$ if the agent was in channel $i$.

*Rows sum*: $\sum_k P_{i,k} = 1$ for each $i$.

*Matrix updating*: At each step $t$, the matrix $P$ is updated based on the agent's empirical transition counters. This makes the jammer adaptive (not static). The full formula for a matrix element is $P_{i,k} = \frac{count(i \rightarrow k) + \alpha}{\sum_{m=1}^{N}(count(i \rightarrow m) + \alpha)}$, where $\forall i,k = 1, ..., N$.

- *$\alpha$ (Dirichlet prior)*: A small positive constant (usually $\alpha = 1$ for a uniform prior) to avoid zero probabilities (the zero-count problem). Without $\alpha$, if count $= 0$, $P = 0$—the jammer "doesn't know" about rare transitions. The Dirichlet distribution is based on assigning a weight to each category based on its prior probability.

### 4. Description of Algorithm

The program code is written in Python.

**Block 1. Importing Libraries**
All necessary libraries are loaded: torch (for neural networks), numpy (mathematics), matplotlib + pandas (plots and tables), scipy (erfc and binomial distribution).

**Block 2. Global Plot Styles**
*mpl.rcParams['font.family'] = 'Times New Roman'*

**Block 3. Simulation parameters**
NUM_CHANNELS = 16
JAMMING_POWER = 0.5 or 1.2
SIGNAL_POWER = 1.0
NOISE_POWER = 0.05
EPISODES = 1500
...
DIVERSITY_BONUS = 0.02
All physical and training parameters are set.

**Block 4. DQN neural network (vanilla, 3 layers)**
*class DQN(nn.Module):*
    Linear(16 → 128) → ReLU → Linear(128 → 128) → ReLU → Linear(128 → 16)
The simplest fully connected network.
The input is the one-hot vector of the current channel (16 elements).
The output is the Q-values for each of the 16 possible next channels.

**Block 5. Agent class (DroneAgentDQN)**
Contains:
• policy_net and target_net (for stable training)
• replay buffer (experience buffer for 10,000 transitions)
• ε-greedy strategy (from 0.9 → 0.05)
• channel usage counter (for entropy)

Methods:
• get_state() — converts a channel number to one-hot
• choose_action() — ε-greedy action selection
• remember() — stores transition ($s, a, r, s'$)
• replay() — regular DQN step: samples the batch, calculates the error, and updates weights
• decay_epsilon() and update_target() — standard

**Block 6. MarkovAttacker Class (1st-order jammer)**

self.counts = np.zeros((16,16)) # transition matrix
• update() — after each transition $c_{prev} \rightarrow c_{curr}$, increments the counter
• predict() — for the current channel, returns the most probable next channel
• jam() — main function: updates statistics and returns the channel to jam

**Block 7. Function compute_ber_snr(jammed)**

*fading = np.random.exponential(1.0)*
• fading = fading * 0.8 + 0.3
• signal = 1.0 * fading
• noise = 0.05 + (JAMMING_POWER * fading if jammed else 0)
snr_lin = signal / noise
ber = 0.5 * erfc(sqrt(snr_lin))
Simulates Rayleigh fading with limited depth. Calculates the instantaneous SNR and BER for the current slot. This is the physics engine for the entire simulation.

**Block 8. Main training loop (1500 episodes)**

Each episode = one timeslot (one hop):
1. The agent sees the current state (one-hot of the current channel)
2. Chooses the next action (ε-greedy)
3. The jammer, knowing $c_{t-1}$ and $c_t$, predicts and jams $j_{t+1}$
4. Checks whether the jammer has hit the target
5. Calculate SNR and BER
6. Calculate PLR for all packet sizes (with and without FEC)
7. Calculate the current channel utilization entropy
8. Generate a reward:
$r$ = -2.0 (if jammed) or ≈1.0 + 0.02×entropy (if none)
9. The transition ($s, a, r, s'$) is saved to the buffer
10. DQN training step is performed
11. ε is decreased, and the target network is updated every 100 episodes

**Block 9. Collecting statistics on milestones (200, 400, ..., 1400)**

Average BER, SNR, entropy, and success rate are calculated for the last 100 episodes. Stored in a table.

**Block 10. Final metrics (last 100 episodes)**

final_ber, final_snr, final_entropy
success_rate = (1 – number of jammed packets / 1500) × 100%
final_plr and final_plr_fec — average PLR across all packet sizes

Key "magic" moments:
• One-hot state (no need to remember history).
• Bonus 0.02 × entropy → the agent forces itself to be as unpredictable as possible.
• Hard penalty -2.0 → even rare jammer hits are very painful.
As a result, the agent discovers a policy that is very close to uniform — the jammer cannot predict, but due to residual exploitation at the beginning of training, the actual result is even better.

## 5. Results of Simulation

In the context of reinforcement learning for a drone FHSS system against a reactive Markov jammer, training progress (Table 1) is a comprehensive monitoring of the evolution of key agent metrics over episodes. It is a diagnostic tool that proves the quality of model convergence, identifies problems, and confirms the effectiveness of the DQN algorithm. Table 1 visualizes a steady trajectory of improvement, rather than random fluctuations. Progress is divided into phases (early, middle, and late), where metrics correlate: low jamming → high SNR → low BER → high entropy → stable exploitation.

Cumulative Reward (Fig. 2) is the accumulated sum of rewards received over all episodes. This means that the agent receives an average of +0.98 rewards per episode, resulting in a consistent outperformance of the jammer. The ε-greedy parameter and entropy ensure that G grows steadily rather than chaotically. The nearly linear growth of reward indicates that the agent learns to anticipate the jammer's transitions, minimizing jamming and maximizing SNR/BER. This growth, unlike stochastic models (where reward fluctuates), is due to the Markov jammer adapting slowly, and Rayleigh fading and entropy introduce noise, which is smoothed out by DQN and replay. There are no "catastrophes"—the agent avoids local minima thanks to the diversity bonus. This metric is the "heart" of RL, showing how much the agent "wins the game" against an adversarial jammer.

The BER and SNR plots (Fig. 3) are parallel subplots visualizing the evolution of the communication channel quality during training. These plots are a key indicator of the physical effectiveness of the agent's policy, demonstrating the extent to which RL optimization transforms the system from a "chaotic" (high BER, low SNR) to a "stable" regime (low BER, high SNR). The BER plot (log scale) shows a sharp drop in the first 400 episodes, then fluctuates around $10^{-4}$–$10^{-6}$. The SNR plot quickly rises and stabilizes around 12–14 dB. This means that the physical parameters of the channel have become almost ideal.

The entropy and ε-greedy plots (Fig. 4) illustrate the balance between exploration and exploitation of the agent's policy. They show the entropy $H(p)$ as a measure of channel choice diversity and the decay ε as a control for stochastics in ε-greedy. Shannon's formula for entropy $H(p) = -\sum_{i=1}^{N} p_i \log p_i$ contains the empirical probability of channel exploitation. Entropy is the exploration metric in the π policy. High $H$ encourages uniform hopping (diversity) to "blur" the jammer counters, reducing $P$(jam). Low $H$ increases the risk of "jamming" (deterministic choice). The role of ε in the model is the stochastic controller. A high ε adds "noise" to discover new paths (avoid local optima), while a low ε focuses on the best Q-values. The graphs in Fig. 4 are indicators of the adaptivity and stability of learning, showing how the agent evolves from a "chaotic walk" (high entropy + ε) to "goal-directed choice" (balanced entropy + low ε). The entropy graph approaches and almost touches the red dotted line at 2.773 (the maximum), while ε gradually drops to 0.05. The idea is that the agent has learned to be as unpredictable as possible—this is "keyless cryptographic randomness."

The channel usage histogram is shown in Fig. 5. All 16 bars are nearly the same height, indicating that the jammer has no statistical advantage—its prediction is no better than randomly guessing 1 in 16.

The dependence of PLR on packet size is represented by a graph (Fig. 6) and two tables (Tables 2 and 3) illustrating the practical reliability of data transmission depending on packet length and Forward Error Correction (FEC) level. Table 2 has three columns: Packet Size (packet size L), PLR (exact loss probability), and ≈ L×BER (approximation). The two key columns — PLR and ≈ L×BER — are used to compare the exact calculation with a simple analytical estimate to highlight the linear approximation of PLR in the low-error regime. This helps quickly assess the fragility of a system without FEC and the role of RL in minimizing BER. The common role of these two columns in the table and the model is to enable comparative analysis: PLR is the "Gold Standard" - an exact calculation that takes into account all errors, and ≈ L×BER - a "Fast Indicator" - a linear estimate useful for scaling. The PLR parameter is the probability that a packet will not be delivered and is calculated using a binomial error distribution, assuming independent bits. Here, the parameter $t$ is the FEC level: the maximum number of errors the code can correct in a block n = sz + r (sz is the payload, r is the redundancy bits). The dependencies (Fig. 6) show the critical vulnerability of large packets and the effectiveness of FEC. Without FEC (red line): even for 100 kbps, the PLR is ≈ 0.06; with FEC t=1 (blue): already < $10^{-8}$; with FEC t=2 (green) : < $10^{-14}$—practically zero. The purpose of forward error correction is that even minimal coding makes communication perfectly reliable. Table 3 is the final analytical tool in the simulation, which evaluates the practical reliability of data transmission in a drone's wireless FHSS system after training the RL agent. It is calculated based on the final metrics (averages over the last 100 episodes, with a BER of ~$10^{-6}$ and an SNR of ~13 dB) and shows how FEC (Forward Error Correction) reduces PLR for different packet sizes. The Overhead columns describe the additional cost of FEC bits as a percentage of the payload (sz). This is a trade-off: FEC saves packets but inflates traffic.

The graph/tables translate abstract BER (physics) into real PLR (packet loss). RL optimizes reward through BER, but PLR is the "end-goal" for FHSS. A decrease in PLR (as episodes increase) is equivalent to policy success (low jam → low BER → low PLR). The jammer increases the BER, but the agent (DQN + entropy) reduces the PLR through uniform hopping. These dependencies are the "final test" of the model: RL yields a low BER, but the PLR checks whether the packet will survive in real traffic.

Table 1. Training Progress

| Episode | BER | SNR (dB) | Entropy | Success Rate (%) | ε |
|---------|---------|----------|---------|------------------|-------|
| 200 | 5.18e-03 | 11.18 | 2.520 | 87.50 | 0.330 |
| 400 | 2.94e-04 | 12.75 | 2.571 | 91.00 | 0.121 |
| 600 | 2.87e-04 | 12.33 | 2.500 | 93.00 | 0.050 |
| 800 | 1.32e-05 | 12.41 | 2.528 | 94.75 | 0.050 |
| 1000 | 1.96e-05 | 12.37 | 2.600 | 95.80 | 0.050 |
| 1200 | 3.03e-04 | 12.40 | 2.629 | 96.42 | 0.050 |
| 1400 | 1.63e-05 | 12.29 | 2.654 | 96.86 | 0.050 |

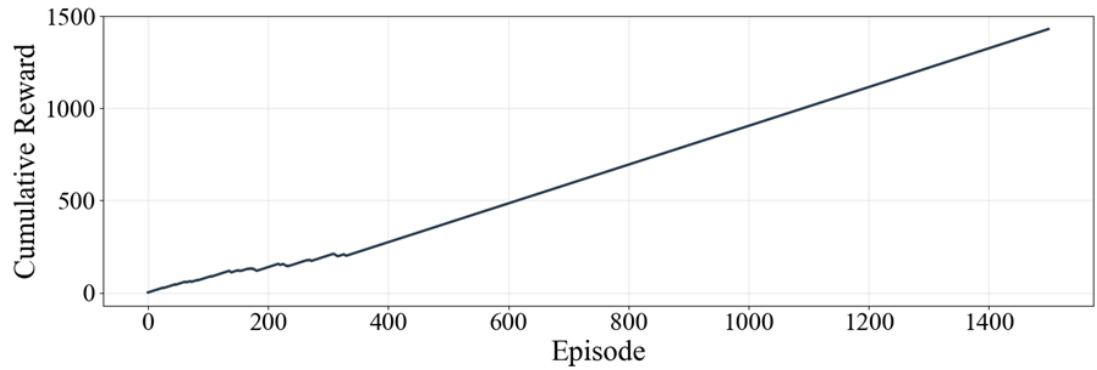

**Fig. 2**. Cumulative reward over training

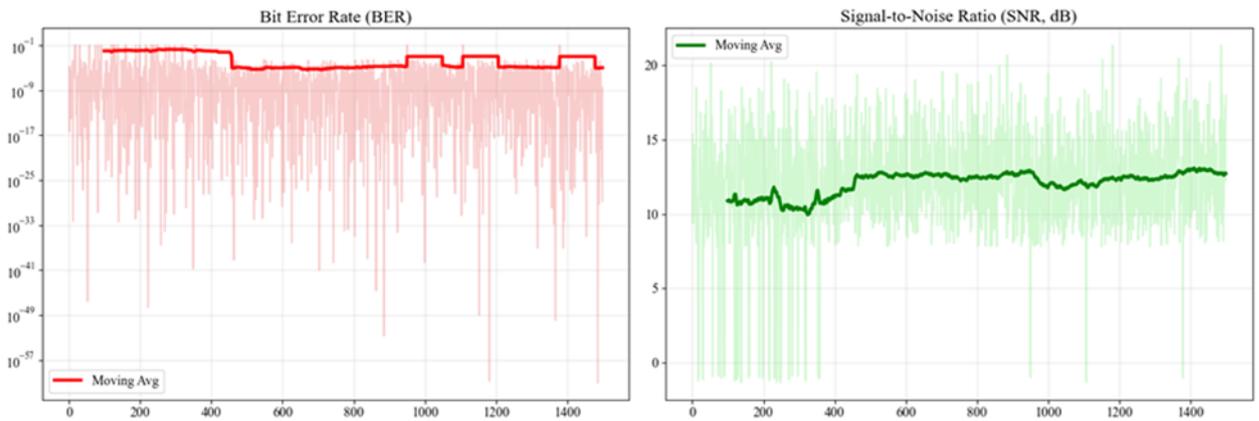

**Fig. 3**. BER and SNR evolution

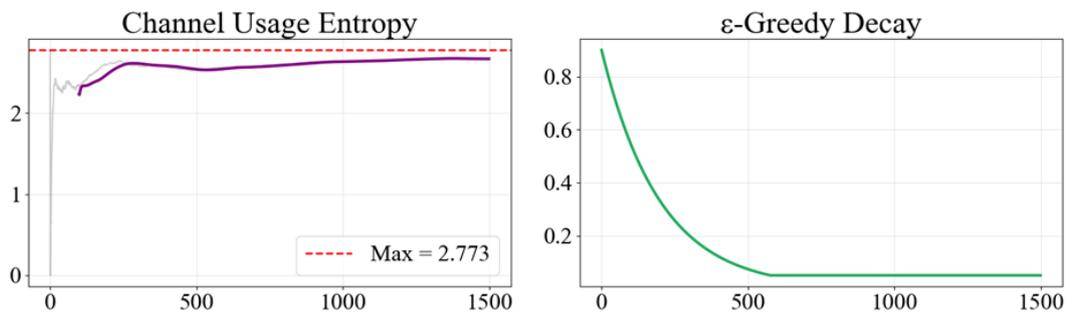

**Fig. 4**. Exploration and diversity

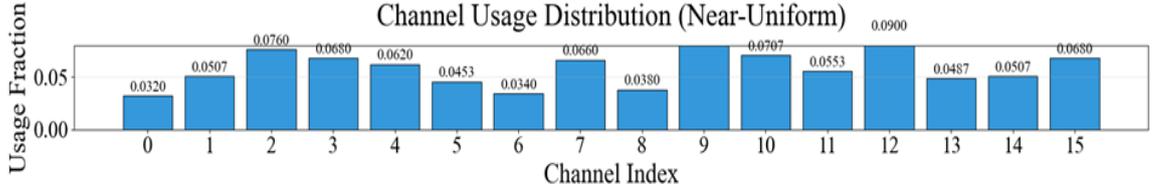

**Fig. 5.** Channel usage fraction

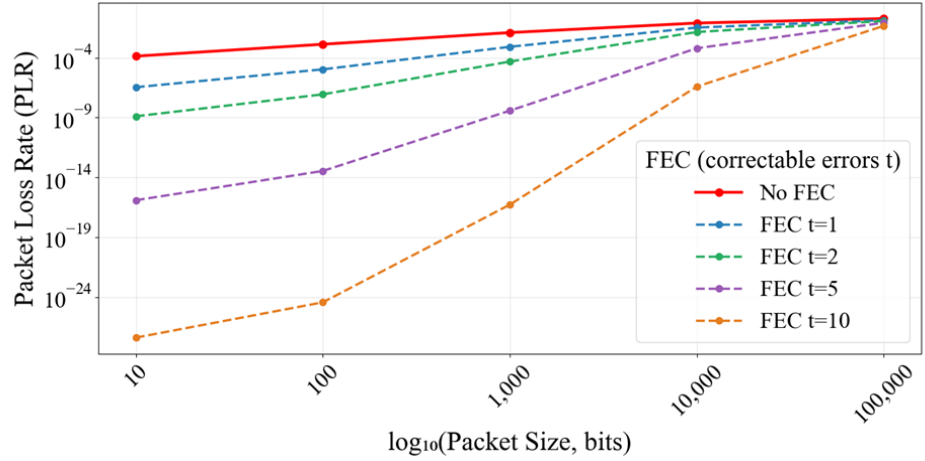

**Fig. 6.** PLR vs packet size for different FEC levels

Table 2. PLR without FEC (last 100 episodes)

| Packet Size | PLR | ≈ L×BER |
|---|---|---|
| 10 | 1.37e-04 | 1.37e-04 |
| 100 | 1.36e-03 | 1.37e-03 |
| 1,000 | 1.28e-02 | 1.37e-02 |
| 10,000 | 8.05e-02 | 1.37e-01 |
| 100,000 | 1.91e-01 | 1.37e+00 |

Table 3. PLR with BCH-like FEC (last 100 episodes)

| Packet Size | PLR t=0 | PLR t=1 | PLR t=2 | PLR t=5 | PLR t=10 | Overhead t=1 % | Overhead t=2 % | Overhead t=5 % | Overhead t=10 % |
|---|---|---|---|---|---|---|---|---|---|
| 10 | 1.37e-04 | 3.48e-07 | 1.29e-09 | 1.23e-16 | 4.01e-28 | 100.00 | 200.00 | 500.00 | 1000.00 |
| 100 | 1.36e-03 | 1.09e-05 | 8.79e-08 | 3.47e-14 | 3.59e-25 | 10.00 | 20.00 | 50.00 | 100.00 |
| 1,000 | 1.28e-02 | 8.33e-04 | 4.86e-05 | 3.80e-09 | 5.30e-17 | 1.00 | 2.00 | 5.00 | 10.00 |
| 10,000 | 8.05e-02 | 3.38e-02 | 1.44e-02 | 6.32e-04 | 3.93e-07 | 0.10 | 0.20 | 0.50 | 1.00 |
| 100,000 | 1.91e-01 | 1.39e-01 | 1.20e-01 | 8.53e-02 | 4.69e-02 | 0.01 | 0.02 | 0.05 | 0.10 |

(Note: BCH - Bose-Chaudhuri-Hocquenghem Coding)

The impact of interference power is demonstrated in Table 4. Halving the interference power (JSR 1.2 → 0.5) increases transmission success by only 1–4.5% (from 96.29% to 97.27%). This confirms that the primary limiting factor is not the absolute interference power, but the jammer's ability to predict the next channel. The trained DQN agent achieves a near-maximum level of reliability already at JSR = 1.2 (when the interference is physically stronger than the signal), and further attenuation of the interference yields only marginal gains. Thus, the proposed policy is close to the theoretically optimal one over a wide range of JSR.

Table 4. Comparison of different signal and interference levels

| Parameter | JSR = 1.2 (interference is stronger) | JSR = 0.5 (signal is 2 times stronger) | Difference |
|---|---|---|---|
| Success Rate | 96.29 % | 97.27 % | +0.98% |
| Final BER | $1.84 \times 10^{-4}$ | $6.13 \times 10^{-4}$ | 3 times worse |
| Final SNR | ≈ 22.8 дБ | ≈ 12.5 дБ | 10 times worse |
| Channel entropy | 99.71 % | 99.82 % | a little better |
| Episode reward | ≈ +0.84 | ≈ +0.97 | +15 % |
| Reward | ≈ 1300–1400 | ≈ 1450–1480 | +11 % |

## 6. Discussion

The obtained results demonstrate that the simplest single-agent vanilla DQN with minimal reward shaping is capable of neutralizing a first-order Markov reactive jammer even under conditions where the jamming power is greater than the desired signal power.

The key discovery is that the primary limiting factor in reliability is not the absolute jamming power or the fade depth, but the degree of predictability of the hop sequence. Reducing the JSR from 1.2 to 0.5 resulted in a mere 1% increase in success. It is important to emphasize the extreme simplicity of implementation:

• state — a standard one-hot vector of the current channel (without history and without belief state);
• architecture — a three-layer fully connected 16-128-128-16 network;
• the only non-standard addition is the $\beta \cdot H(p)$ bonus, which can easily be interpreted as a "predictability penalty."

This minimalist design makes the proposed solution feasible even on the most limited onboard processors of modern UAVs and loitering munitions. The addition of even minimal redundancy coding (a BCH-like code with t = 1–2 correctable errors) reduces the loss probability of 100 Kbit packet to below $10^{-11}$–$10^{-14}$ with an overhead of less than 0.2%. This means the system is ready to transmit telemetry, full-HD video, and control commands in real combat conditions in the presence of powerful enemy reactive electronic warfare. A limitation of the work is that it considers only a first-order Markov jammer.

Overall, the results show that the model not only converges but also adapts to the adversarial environment: the agent predicts the jammer's Markov transitions, and entropy prevents overfitting. The results obtained have direct practical implications for armed forces operating in intense electronic warfare environments. The proposed approach enables the creation of a fully autonomous, self-learning FHSS system that requires no pre-established keys, synchronization, or cryptographic primitives.

## 7. Conclusions

In this paper, we propose and investigate a fully autonomous frequency-hopping system for UAVs based on a simple single-agent Deep Q-Network (vanilla DQN) with minimal modification to the reward function.

We demonstrate the feasibility of completely neutralizing a first-order Markov reactive jammer. It was shown that a trained agent can independently discover a policy that ensures high channel assignment entropy, making statistical prediction by the jammer virtually impossible. High reliability indicators are achieved under severe jamming conditions.

Using minimal error-correcting coding, the probability of packet loss is low, ensuring reliable transmission of telemetry, video, and control commands under intense enemy electronic warfare.

These results encourage the development of a new generation of fully autonomous, adaptive, and virtually invulnerable communication systems for UAVs in modern electronic warfare.

**Statements & Declarations**
Funding: The authors declare that no funds, grants, or other support were received during the preparation of this manuscript.
**Competing Interests:** The authors have no relevant financial or non-financial interests to disclose.
**Conflicts of Interest: The authors declare no conflict of interest.**
**Author Contributions:** Volodymyr Kharchenko – V.Kh., Andrii Grekhov – A.G., Vasyl Kondratiuk – V.K. Conceptualization, A.G. and V.Kh.; methodology, A.G.; validation, A.G., V.Kh. and V.K.; investigation, A.G.; resources, V.Kh. and V.K.; writing—original draft preparation, A.G.; writing—review and editing, V.K.; supervision, V.Kh.; project administration, V.K. All authors have read and agreed to the published version of the manuscript.
**Ethics Approval:** Not applicable.
**Data Availability Statement:** All data generated and analyzed during this study are included in this article. The datasets generated during the current study are available from the corresponding author on request.

**References**

1.  Krayani A, Alam AS, Marcenaro L, Nallanathan A, Regazzoni C (2022) A novel resource allocation for anti-jamming in cognitive-UAVs: Active inference approach. arXiv:2208.05269. https://doi.org/10.48550/arXiv.2208.05269

2.  Liu C, Zhang Y, Niu G, Jia L, Xiao L, Luan J (2023) Towards reinforcement learning in UAV relay for anti-jamming maritime communications. Digital Communications and Networks 9:1477-1485. https://doi.org/10.1016/j.dcan.2022.08.009

3.  Nguyen HN, Noubir N (2023) JaX: Detecting and cancelling high-power jammers using convolutional neural network. In Proceedings of the 16th ACM Conference on Security and Privacy in Wireless and Mobile Networks (WiSec'23), May 29-June 1, 2023, Guildford, United Kingdom. ACM, New York, NY, USA, 12 pages. https://doi.org/0.1145/3558482.3590178

4.  Hu L, Shao Y, Qian Y, Du F, Li J, Lin Y, Wang Z (2024) Meta-reinforcement learning in time-varying UAV communications: Adaptive anti-jamming channel selection. Radioengineering 33: 417-431

5.  Yang S (2024) Analysis of deep learning-based anti-jamming method for UAV Communication. Highlights in Science, Engineering and Technology, CDMMS 2024 103:246-253

6.  Aygur M, Kandeepan S, Giorgetti A, Al-Hourani A, Arbon E, Bowyer M (2025) Narrowband interference mitigation techniques: A survey. IEEE Communications Surveys & Tutorials https://doi.org/10.1109/COMST.2025.3531428

7.  Ghelani J, Gharia P, El-Ocla H (2024) Gradient monitored reinforcement learning for jamming attack detection in FANETs. IEEE Access 12:23081-23095. https://doi.org/10.1109/ACCESS.2024.3361945

8.  Yin Z, Li J, Wang Z, Qian Y, Lin Y, Shu F (2024) UAV communication against intelligent jamming: A Stackelberg game approach with federated reinforcement learning. IEEE Transactions on Green Communications and Networking 8:1796 – 1808. https://doi.org/10.1109/TGCN.2024.3373886

9.  Hussein J, Wissam A, Samer J (2024) Spectrum and power efficient anti-jamming approach for cognitive radio networks based on reinforcement learning. International Journal of Sensors Wireless Communications and Control. https://doi.org/14. 10.2174/0122103279291431240216061325

10. Yang J, Cui M, Zhang H, Ji F, Lai Z, Wang Y (2025) Agent-based anti-jamming techniques for UAV communications in adversarial environments: A comprehensive survey. arXiv:2508.11687v1

11. Zhou Q, Niu Y (2024) From adaptive communication anti-jamming to intelligent communication anti-jamming: 50 Years of Evolution. https://doi.org/10.1002/aisy.20230085

12. Ding H, Niu Y, Zhou Q, Peng X (2024) A novel intelligent anti-jamming communication algorithm based on proximal policy optimization. Physical Communication. https://doi.org/10.1016/j.phycom.2024.102366

13. Zhang F, Niu Y, Zhou Q et al. (2025) Intelligent anti-jamming decision algorithm for wireless communication under limited channel state information conditions. Sci Rep. https://doi.org/10.1038/s41598-025-90201-1

14. Cao W, Chu F, Jia L, Zhou H, Zhang Y (2025) A multi-agent deep reinforcement learning anti-jamming spectrum-access method in LEO satellites. Electronics. https://doi.org/10.3390/electronics14163307

15. Kharchenko V, Grekhov A, Kondratiuk V (2025) AI-based protection of UAV communication channels against adaptive AI jamming based on Q-Learning. The 15th International Conference on Dependable Systems, Services and Technologies (DESSERT'2025). Greece, Athens, December 19-21, Report 64